\documentclass[amsmath,amssymb,twocolumn,pra,showpacs]{revtex4-1}

\usepackage{graphicx,comment,color}
\usepackage{epstopdf}
\usepackage{dcolumn}
\usepackage{bm}
\usepackage[T1]{fontenc}
\usepackage[latin1]{inputenc}
\usepackage{enumerate}
\usepackage{fancyhdr,color}
\usepackage{verbatim}
\usepackage{hyperref}
\usepackage{mathtools}

\def\cH{\mathcal{H}}

\def\cL{\mathcal{L}}

\def\cO{\mathcal{O}}

\def\Tr{\mathrm{Tr}}
\def\Pr{\mathrm{Pr}}

\newcommand{\ket}[1]{\left| #1\right\rangle}        
\newcommand{\sket}[1]{| #1\rangle}  
\newcommand{\bra}[1]{\left\langle #1\right|}        

\begin{document}

\title{Quantum circuit synthesis for generalized coherent states}

\author{
Rolando D. Somma}
\affiliation{Theoretical Division, Los Alamos National Laboratory, Los Alamos, NM 87545, US.}

\date{\today}

 \begin{abstract}
 We present a method that outputs a sequence of simple unitary
 operations to prepare a
 given  quantum state that is a generalized
 coherent state. Our method takes as inputs
 the expectation values of some relevant observables
 on the state to be prepared. Such expectation values
 can be estimated by performing projective measurements
 on $\cO(M^3 \log(M/\delta)/\epsilon^2)$ copies of the state, where $M$ is the dimension
 of an associated Lie algebra, $\epsilon$ is a precision parameter, 
 and $1-\delta$
 is the required confidence level. The method 
 can be implemented on a classical
 computer and runs in time $\cO(M^4 \log(M/\epsilon))$.
 It provides $\cO(M \log(M/\epsilon))$ 
 simple unitaries
 that form the sequence. The number of all computational resources
 is then polynomial in $M$, making the whole procedure very efficient in those
 cases where $M$ is significantly smaller than the Hilbert space dimension.
 When the algebra of relevant observables
 is determined by some Pauli matrices,
 each simple unitary may be easily decomposed into 
  two-qubit gates.
 We discuss applications to quantum state tomography
 and classical simulations of quantum circuits.
  \end{abstract}

\maketitle          

\section{Introduction}
\label{sec:intro}

An important problem
in quantum information theory 
regards the determination of a pure multiparty
quantum state $\ket \psi$.
This problem has many applications, including
quantum device and quantum state verification.
The most general procedure to determine $\ket \psi$
is via quantum state tomography~\cite{DPS03}. In this case,
a complete set of observables must be measured
on multiple copies of the state. The number of measurements
depends then on the number of observables as well as the
precision and confidence levels required in the state reconstruction.
This number is prohibitively large in general, i.e.,
it is polynomial in the Hilbert space dimension
or exponential in the size of the quantum system~\cite{Hol73}.

For this reason, some recent approaches focus
on cases where a form of quantum state
tomography can be performed efficiently~(c.f., \cite{AG08,CPF+10,Mon17}). 
These approaches
work only for particular classes of states and do not 
apply generally.
Also of interest are quantum tomographic
approaches for more general pure quantum states
that can be prepared by a quantum computer via a quantum circuit.
This problem may be important for the validation and
verification of quantum computing devices,
which is also studied
in Refs.~\cite{BFK08,FK17,Mah18} from a different viewpoint.
In this problem, the assumption is that the length of the quantum
circuit is significantly (exponentially)
smaller than the Hilbert space dimension.
Then, an efficient description for the prepared
quantum state $\ket \psi$ is given by
the unitary operators
that form the circuit.  
That is,
quantum state tomography of $\ket \psi$ is also achieved by providing
a quantum circuit that prepares it. As in general quantum
state tomography, the quantum circuit
may be obtained by measuring certain observables
on multiple copies of $\ket \psi$. Once the quantum circuit is known,
additional information about the state may be obtained efficiently
using classical resources.

Some known results already
use this ``quantum circuit'' approach to quantum state tomography.
In Ref.~\cite{CPF+10}, for example, an efficient method for quantum state
tomography of matrix product states is given. It is shown
that, by obtaining the reduced density matrices of the $n$-qubit
quantum state $\ket \psi$, a quantum circuit of  size polynomial in $n$ that
prepares $\ket \psi$ can be constructed. In Ref.~\cite{Mon17}, a similar
approach is analyzed for the so-called stabilizer states.
That approach is based on performing Bell measurements
to determine a complete set of generators for the stabilizer
operators. Using the results of Ref.~\cite{AG08}, a quantum circuit 
of polynomial size that prepares the stabilizer state can be obtained.
We note that in these examples,
knowing a quantum circuit that prepares the state
also allows for the efficient classical computation of expectation
values of other observables on $\ket \psi$.

Here, we analyze this approach
in the case of generalized coherent states (GCSs).
We call our approach quantum circuit synthesis for GCSs
as the problem reduces to finding a sequence of simple unitary
transformations that prepare a given GCS, when acting on some trivial
initial state. When dealing with $n$-qubit systems, each
unitary transformation may be easily decomposed into a polynomially large
sequence of two-qubit gates, in the worst case, and the initial state may be $\ket{00\ldots0}$. When dealing with quantum systems obeying
different particle statistics, such as fermionic
or bosonic systems, the results in Refs.~\cite{SOGKL02,SOK03}
may be useful for simulating such systems on quantum computers
and devising quantum circuits that implement the unitary transformations.

GCSs have the nice property that are uniquely determined by the expectation
values of certain observables.  These observables form a basis of
an associated Lie algebra. For many important cases, the number of observables 
in the basis is significantly (exponentially) smaller than the Hilbert
space dimension. It is then reasonable
to consider and develop efficient methods that take the expectation
values as input and provide the sequence
of unitary transformations that prepare the GCS as output. 
We provide one such method in this paper.

Our main result can be interpreted as
an efficient quantum-circuit approach to quantum state tomography
of GCSs. The number of copies of the state
needed to estimate the expectation values of observables
and the number of classical
operations to find the unitaries (or quantum circuit) scales
polynomially with $M$, $1/\epsilon$, and $\log(1-\delta)$. Here,
$M$ is the dimension of the associated Lie algebra, $\epsilon>0$
is a precision parameter, and $\delta<1$ is the confidence level. 
The latter appears naturally
from the estimation of the expectation values by repeated measurements.
Our method relies heavily on a {\em diagonalization} procedure 
that is suitable for Lie algebras~\cite{Som05,Wil93},
and which is a generalization of the Jacobi method to diagonalize
matrices.

Additional implications of our main result are in order.
First, the results in Ref.~\cite{SBO06} can be used
to obtain expectation values of certain unitary operations
and other observables on the GCS efficiently,
using only classical resources. Such observables
do not necessarily belong to the Lie algebra.
Second, our techniques can be applied to simulate
certain classes of quantum circuits efficiently on a classical computer.

Our paper is organized as follows.
In Sec.~\ref{sec:GCS} we introduce 
 GCSs and give some results about
semisimple Lie algebras.
In Sec.~\ref{sec:problem} we formalize 
the problem that we are solving. In Sec.~\ref{sec:circuit}
we provide the details of our method, 
describe the diagonalization procedure,
and obtain bounds for the number of copies of the state
and classical operations required.  
In Sec.~\ref{sec:applications}
we discuss further applications of our results
to the efficient classical simulation of
certain quantum circuits.

\section{Generalized coherent states}
\label{sec:GCS}
To introduce GCSs~\cite{Gil74,Per85,ZFG90}, 
we focus on the case where a set 
of observables (Hermitian operators) form a semisimple Lie algebra 
$\mathfrak h$ of linear operators
acting on a finite dimensional
Hilbert space $\cH$.
The Hilbert space
can be that of a multiparty quantum system.
We assume $\mathfrak h$ to be a real Lie
algebra of dimension $M$. The bracket of two observables
$O_x$ and $O_y$ is given by $[O_x,O_y]=i (O_xO_y-O_yO_x)$,
which is familiar to physicists.
We also assume that $\mathfrak h$ acts irreducibly on $\cH$.

If $O_x \in \mathfrak h$, the Lie group
generated by $\mathfrak h$ involves the map
$O_x \rightarrow e^{iO_x}$. GCSs associated with $\mathfrak h$ are the pure quantum
states
\begin{align}
\label{eq:GCS}
   \{ \ket \psi \equiv e^{i \mathfrak h}\ket{\rm{hw}}\} \;,
\end{align}
where $e^{i \mathfrak h}$ is any unitary Lie group
operation, commonly referred to as a displacement operator.
$\ket{\rm{hw}}$ is a pure quantum state
that represents the highest-weight state of $\mathfrak h$, as explained below.
That is, GCSs belong to the unique orbit of a highest-weight
state of $\mathfrak h$ under the action of the Lie group.

As $\mathfrak h$ is semisimple,
it assumes a Cartan-Weyl basis decomposition:
$\mathfrak h_D \oplus \mathfrak h^+ \oplus \mathfrak h^- $~\cite{Hum72,Geo99,Hal03}.
The subalgebra $\mathfrak h_D$ is the Cartan subalgebra (CSA)
  and is spanned by the largest set of commuting
observables in $\mathfrak h$. The generators of the CSA are $H_1,\ldots,H_R$.
$\mathfrak h^+$
  is spanned by $L$ raising 
operators $E^+_{1},\ldots, E^+_L$ and 
$\mathfrak h^-$
  is spanned by $L$ lowering 
operators $E^-_{1}=(E^+_1)^\dagger,\ldots, E^-_L=(E^+_L)^\dagger$.
The weight states of $\mathfrak h$ are simultaneously eigenstates of 
all observables in $\mathfrak h_D$. In particular, given a Cartan-Weyl basis
for $\mathfrak h$,
the highest-weight state $\ket{\rm{hw}}$
is a weight state that satisfies
\begin{align}
\label{eq:raising}
    E^+_l \ket{\rm{hw}} = 0 \; , \; \forall \; 1 \le l \le L \;.
\end{align}
The {\emph {weights}} of $\ket{\rm{hw}}$, $w:\mathfrak h_D \rightarrow \mathbb R$, are defined by $H_r\ket{\rm{hw}}=w(H_r)\ket{\rm{hw}}$.

The GCSs given by Eq.~\eqref{eq:GCS} should be compared
with the well-known Glauber coherent states appearing in quantum optics~\cite{Gla63}.
That is, for coherent states, the vacuum state plays the role of
the highest-weight state
and the displacement operator is given by 
the exponential of field operators.
GCSs are then a generalization of coherent states 
that may be suitable
for finite-dimensional quantum systems.

\section{Statement of the problem}
\label{sec:problem}
\begin{figure}[h]
    \centering
  \includegraphics[width=5cm]{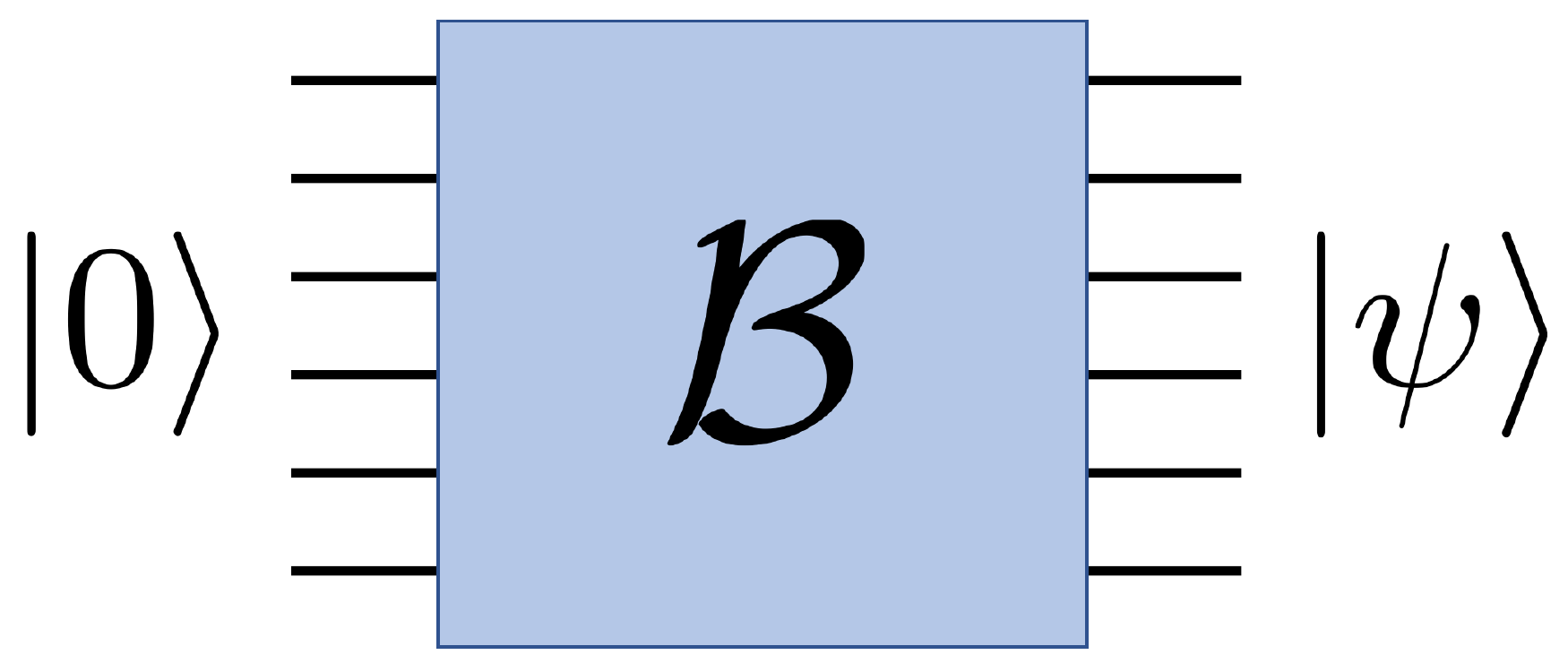}
  \caption{Black-box GCS preparation. The initial state
  $\ket 0$ may represent the highest-weight state and
  $\cal B$ is a quantum operation that transforms $\ket 0$
  into the GCS $\ket \psi$.}
  \label{fig:BlackBox}
\end{figure}

We let $\ket \psi$ be a GCS 
associated with the Lie algebra $\mathfrak h$
and $\mathfrak h_D \oplus \mathfrak h^+ \oplus \mathfrak h^-$
denotes a given Cartan-Weyl basis.
$\ket \psi$ is prepared using a black-box
transformation as in Fig.~\ref{fig:BlackBox}, and we assume that 
we have access to multiple copies of $\ket \psi$.
The goal
is to obtain a sequence of {\em simple} unitary transformations $U_1,\ldots,U_K$
such that, with probability at least $1-\delta$,
\begin{align}
\label{eq:ProcedureError}
    \| \ket \psi - \prod_{k=1}^K U_k \ket{\rm{hw}} \| \le \epsilon \;.
\end{align}
$\delta >0$ and $\epsilon>0$ are given
and determine the confidence level (i.e., $1-\delta$) and precision of the estimation, respectively.
The unitaries $U_k$ belong to the Lie group obtained
from $\mathfrak h$ and, for the given Cartan-Weyl basis, they are
\begin{align}
\label{eq:GroupOps}
    U_k = \exp \left \{ i \left(\alpha_k E^+_{l(k)} + \alpha^*_k E^-_{l(k)} \right) \right \} \;,
\end{align}
where $\alpha_k \in \mathbb C$ and $ l(k) \in \{1,\ldots,L \}$.

For qubit systems, the Lie algebra $\mathfrak h$
is usually spanned by certain
products of Pauli operators acting on qubit spaces.
In this case, the unitaries $U_k$ may be easily decomposed
as sequences of one and two-qubit gates following, for example,
the results in Ref.~\cite{SOGKL02}. For some relevant
Lie algebras, such as $\mathfrak{so}(2n)$, the number of gates in this sequence
scales linearly with the number of qubits, in the worst case~\cite{SOB04}.

\section{Unitary operations for GCS\scriptsize{s}}
\label{sec:circuit}
Any GCS $\ket \psi$ is the unique ground state (eigenstate of lowest
eigenvalue) of 
some observable in $\mathfrak h$~\cite{BKO03}.
It follows that, if $\{O_1,\ldots,O_M\}$ is a basis of observables for $\mathfrak h$, $\ket \psi$ is uniquely determined (up to an irrelevant
phase factor)
by the expectation values $\langle O_m \rangle_\psi = \bra \psi O_m \ket \psi$,
$1 \le m \le M$~\cite{SCB06}.
To obtain the sequence of unitaries $U_k$ of Eq.~\eqref{eq:ProcedureError}
that prepares a given GCS, or a quantum circuit,
the first goal is then 
to determine the expectation values. 
This will require performing (projective) measurements of each $O_m$
in multiple copies of $\ket \psi$. 
The number of copies will depend on the goal precision
and confidence level; however, we assume first
that the expectations can be obtained exactly
and analyze the effects of errors in Sec.~\ref{sec:prec-conf}.

\begin{figure}[h]
    \centering
  \includegraphics[width=8.7cm]{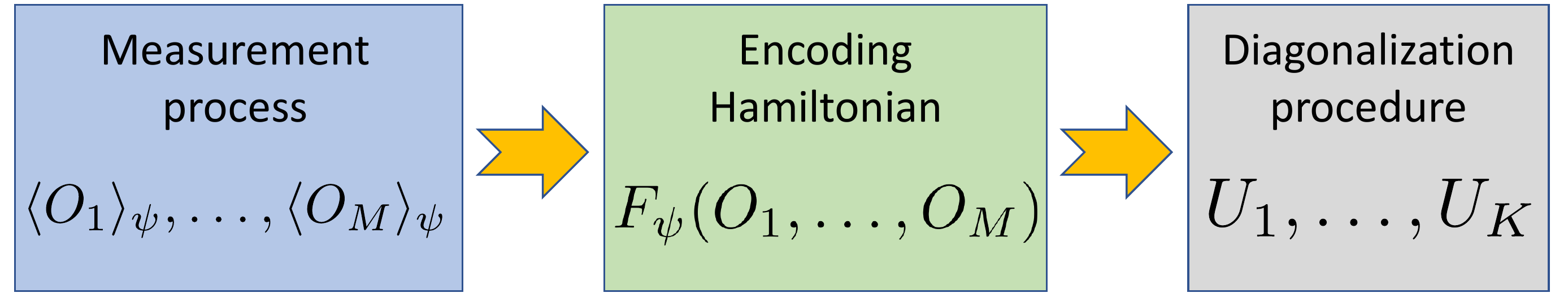}
  \caption{Our approach to obtaining a sequence of unitary transformations
  $U_1,\ldots,U_K$ to prepare a GCS, as determined by Eq.~\eqref{eq:ProcedureError}.}
  \label{fig:Procedure}
\end{figure}

Our procedure to find the $U_k$'s is as follows.
From the expectation values $\langle O_m \rangle_\psi$,
we will construct a Hamiltonian (i.e., an observable in $\mathfrak h$)
that has $\ket \psi$ as unique eigenstate of largest eigenvalue.
Then, we will invoke a classical procedure to \emph{diagonalize} the Hamiltonian
using Lie group operations. Such a procedure was used in Refs.~\cite{Som05,SBO06}
in the context of \emph{generalized mean-field Hamiltonians}
and is based on an extension of Jacobi's diagonalization method
that applies to semisimple Lie algebras~\cite{Wil93}.
The Lie group operations
will correspond to the $U_k$'s.
Figure~\ref{fig:Procedure} summarizes our approach.

If the basis of observables for $\mathfrak h$
is orthogonal, it satisfies~\cite{SOGKL02}:
\begin{align}
\label{eq:hinnerproduct}
    \Tr{\left(O_mO_{m'}\right)}={\cal N}. \delta_{mm'}\;,
\end{align}
with ${\cal N}>0$.
The orthogonal basis may be obtained from
the given Cartan-Weyl basis, i.e.
\begin{align}
\label{eq:CWbasis}
\begin{cases}
     & O_1=H_1,\ldots, O_R=H_R \;, \\
     &  O_{R+1}=(E^+_1+E^-_1),\ldots,O_{R+L}=(E^+_L+E^-_L), \\
      & O_{R+L+1}=i(E^-_1-E^+_1),
\ldots, O_{R+2L}=i(E^-_L-E^+_L).
\end{cases}
\end{align}
A preprocessing step may be needed to satisfy
Eqs.~\eqref{eq:hinnerproduct} and~\eqref{eq:CWbasis},
but that step is not very time consuming -- it may
require multiplying the operators in the Cartan-Weyl basis by
different constants -- and will not dominate the complexity
of our procedure.
Then, the expectation values $\langle O_m \rangle_\psi$
are determined from the expectation values of the operators
in the given Cartan-Weyl basis (and vice versa).
We will assume that Eq.~\eqref{eq:CWbasis} is satisfied.

Given the relevant expectation values, we construct
the following Hamiltonian in $\mathfrak h$:
\begin{align}
    F_\psi = \sum_{m=1}^M  O_m \langle O_m \rangle_\psi \;.
\end{align}
For any pure quantum state $\ket \phi \in \mathcal H$, it is known that
\begin{align}
   0 \le \sum_{m=1}^M  \langle O_m \rangle^2_\phi \le P_{\mathfrak h} \; ,
\end{align}
where $P_{\mathfrak h}>0$ is the so-called purity relative to the
algebra $\mathfrak h$ (or $\mathfrak h$-purity). Furthermore, for any GCS $\ket \psi$ associated with $\mathfrak h$, we have
\begin{align}
\label{eq:hpurity}
    \sum_{m=1}^M  \langle O_m \rangle^2_\psi = P_{\mathfrak h}  \; .
\end{align}
The $\mathfrak h$-purity is then invariant under Lie group transformations.
Quantum states that are not GCSs have $\mathfrak h$-purity strictly less than $P_{\mathfrak h}$-- see Ref.~\cite{BKO03} for more details.

The Cauchy-Schwarz inequality implies
\begin{align}
  \left|  \langle F_\psi \rangle_\phi \right|= \left|\sum_{m=1}^M \langle O_m \rangle_\phi \langle O_m \rangle_\psi \right| \le P_{\mathfrak h} \; .
\end{align}
Furthermore, the upper bound is tight
and is only obtained when $\ket \phi= \ket \psi$,
since $\langle F_\psi \rangle_\psi$ reduces to the
$\mathfrak h$-purity defined in Eq.~\eqref{eq:hpurity}.
As GCSs are uniquely determined by the
expectation values of the basis operators,
the state $\ket \psi$ is then the unique eigenstate
of $F_\psi$ of largest eigenvalue $P_{\mathfrak h}$.

We let $U$ be the Lie group operation
that transforms  $\ket{\rm{hw}}$ into the GCS
$\ket \psi$ as in Eq.~\eqref{eq:GCS}.
We also define the Hamiltonian $F_{\rm {hw}} = \sum_{r=1}^{R} w(H_r) H_r$,
where $w(H_r) \in \mathbb R$ are the eigenvalues of $\ket{\rm{hw}}$
associated with each $H_r$ (weights). 
In the following, we will show $U^\dagger F_{\psi} U =F_{\rm {hw}}$. That is,
$U$ can be used to {\em diagonalize}
$F_\psi$. By diagonalization we mean a particular transformation that
takes an element of $\mathfrak h$ and maps it into an observable in
$\mathfrak h_D$.

From Eq.~\eqref{eq:raising}, the raising and lowering operators satisfy
$\langle E^{\pm}_l \rangle_{\rm hw}=0$ for all $l$.
Then, Eq.~\eqref{eq:CWbasis} implies $F_{\rm hw}=\sum_{m=1}^M \langle O_m \rangle_{\rm hw} O_m$.
The unique eigenstate of $F_{\rm {hw}}$
of largest eigenvalue $P_{\mathfrak h}$ is then $\ket{\rm{hw}}$.
Since $U$ is a unitary transformation in the Lie group, we obtain
\begin{align}
    U F_{\rm{hw}} U^\dagger = \sum_{m=1}^M c_m O_m \;,
\end{align}
where $c_m \in \mathbb R$. We can obtain the coefficients using Eq.~\eqref{eq:hinnerproduct} as follows:
\begin{align}
\label{eq:coeff1}
    c_m = \Tr{\left(U F_{\rm hw} U^\dagger O_m \right)}/{\cal N} \;.
\end{align}
With no loss of generality, $U^\dagger O_m U = \sum_{m'=1}^M d_{mm'}O_{m'}$,
where $d_{mm'} \in \mathbb R$.
Then, Eq.~\eqref{eq:coeff1} can be rewritten as
\begin{align}
\label{eq:coeff2}
    c_m = \sum_{r=1}^{R}   d_{mr} w(H_r) \;.
\end{align}
Similarly,
\begin{align}
\nonumber
    \bra{\rm{hw}} U^\dagger O_m U \ket{\rm{hw}} &= \sum_{m'=1}^{M}d_{mm'}
     \bra{\rm{hw}}   O_{m'}   \ket{\rm{hw}} \\
     \label{eq:coeff3}
      & = \sum_{r=1}^{R} d_{mr} w(H_r) \;.
\end{align}
As Eqs.~\eqref{eq:coeff2} and~\eqref{eq:coeff3} coincide, 
we obtain $c_m = \langle O_m \rangle_\psi$ and
\begin{align}
\label{eq:transfH}
    F_{\rm hw}= U^\dagger F_\psi  U  \;.
\end{align}

We now assume there exists a classical procedure to
diagonalize $F_\psi$ that outputs a sequence of
Lie-group unitaries $V_1,V_2,\ldots,$ that form a unitary
$V$. We also let each $V_k$ 
be an exponential like the one in Eq.~\eqref{eq:GroupOps}.
By the assumption, $V^\dagger F_\psi V$ is
an element of the CSA; we define it as $F_D$.
The largest eigenvalue of $F_D$ is also $P_{\mathfrak h}$
and the corresponding eigenvector is the weight state
$\ket{\rm w}$.

In principle, $F_D \ne F_{\rm hw}$. The equality
would only hold if $\ket{\rm w} = \ket{\rm hw}$.
Nevertheless, since both $V$ and $U$ are in the Lie group, there
is a Lie group transformation that takes $\ket{\rm w} \rightarrow \ket{\rm hw}$ and thus $F_D \rightarrow F_{\rm hw}$.
This last transformation can be built upon a sequence of Weyl-group
reflections. The sequence
can be determined efficiently in $M$ and
each reflection also takes the form of Eq.~\eqref{eq:GroupOps}--
see Thm. 2.63 in Ref.~\cite{Kna96}.

\subsection{Diagonalization procedure}
\label{sec:DP}
A sequence of
unitaries having the form of Eq.~\eqref{eq:GroupOps} can then be obtained via a classical procedure that diagonalizes $F_\psi$. This procedure
is discussed in Sec. 5.2.1 of Ref.~\cite{Som05} and is based on
Ref.~\cite{Wil93}. It can be interpreted as a generalization
of Jacobi's diagonalization method suitable to the case
of Lie algebras. While in this section we analyze the 
steps to diagonalize $F_\psi$, the same procedure can be used
to diagonalize any Hamiltonian in $\mathfrak h$.
We summarize the procedure here for completeness,
while still providing more details than Ref.~\cite{Som05}.
The actual complexity of this procedure for our problem
is determined in Sec.~\ref{sec:prec-conf}.

As before, we assume that the expectations
$\langle O_m \rangle_\psi$ are exactly known
and analyze the effects of errors in these expectations
in Sec.~\ref{sec:prec-conf}. 
To be {\em exact}, the diagonalization procedure should involve 
infinitely many steps. Nevertheless,
the number of steps can be made finite with controlled errors. 
We  define
a sequence of Hamiltonians in $\mathfrak h$, $\{F^0_\psi, F^1_\psi, \ldots\}$, with  $F^0_\psi=F_\psi$. Step $k$ takes $F^{k-1}_\psi$ as input
and outputs $F^{k}_\psi$, which is closer
to being diagonal, according to a measure that provides
the distance between observables in $\mathfrak h$ and the CSA--
see Sec.~\ref{sec:prec-conf}. From Eq.~\eqref{eq:CWbasis},
each $F^k_\psi$ admits the following representation:
\begin{align}
\label{eq:Fdecomp}
    F^k_\psi = \sum_{r=1}^R \gamma^k_r H_r + \sum_{l=1}^L \left(\iota^k_l E^+_l + (\iota^k_l)^* E^-_l \right)\;.
\end{align}
The coefficients $\gamma^0_r$ and $\iota^0_l$ may be obtained
from the initial expectation values 
$\langle O_m \rangle_\psi$.
We need to update and store in memory
the coefficients $\gamma^k_r$ and $\iota^k_l$  at each step.

Before describing each step, we present
some results from the theory of semisimple Lie algebras
that will be useful. For each $l$, the operators $E^+_l$, $E^-_l$, and some observable in the CSA form
an $\mathfrak{su}(2)$ algebra (Thm. 7.19 of~\cite{Hal03}). 
This algebra can be obtained as follows.
First, we compute the commutator $[E^+_l,E^-_l] = Z_l$,
where $Z_l$ is nonzero and $Z_l\in \mathfrak h_D$;
we write $Z_l=\sum_{r=1}^R \mu_{lr} H_r$.
Then, we compute $[E^+_l,Z_l]=\eta_l Z_l$; with no loss
of generality, $\eta_l > 0$ (otherwise
replace $E^+_l$ by $E^-_l$). The $\mathfrak{su}(2)$ algebra
is constructed from $S^+_l = E^+_l/\sqrt{ \eta_l}$, $S^-_l = E^-_l/\sqrt{ \gamma_l}$,
and $S^z_l = Z_l/ \eta_l$. These operators satisfy
the usual $\mathfrak{su}(2)$ commutation relations:
$[S^+_l,S^-_l]=S^z_l$ and $[S^z_l,S^{\pm}_l]=\pm S^{\pm}_l$.
Alternatively, we can define the observables $S^x_l = (S^+_l+S^-_l)/\sqrt 2$
and $S^y_l = i(S^-_l-S^+_l)/\sqrt 2$. In physics,
these observables are associated with the angular momentum
operators.

At the $k$-th step of the diagonalization procedure,
we search for the value of $l'$ such that
$|\iota^{k-1}_{l'}|\ge |\iota^{k-1}_l|$, for all $l \in \{1,\ldots,L\}$.
We then construct a unitary $V_k$ that is of the form
$e^{i (\alpha_k E^+_{l'} + \alpha^*_k E^-_{l'})}$. 
This unitary will be used to transform $F_\psi^{k-1}$ into
$F_\psi^k$.
The coefficients
$\alpha_k$ can be obtained
as follows.
With no loss of generality, 
\begin{align}
\label{eq:kHamildecomp}
    F^{k-1}_\psi= \xi^z S^z_{l'}  
    + \xi^x S^x_{l'}+ \xi^y S^y_{l'} + X^\perp_{l'} \;,
\end{align}
where $\xi^z, \xi^x, \xi^y \in \mathbb R$ and $X^\perp_{l'} \in \mathfrak h$
is a Hamiltonian that is orthogonal to the corresponding $\mathfrak{su}(2)$ algebra,
according to the inner product of Eq.~\eqref{eq:hinnerproduct}.
The constants satisfy
\begin{align}
\label{eq:xix}
    \xi^x &= \sqrt{\eta_{l'}/2}\left(\iota^{k-1}_{l'}+(\iota^{k-1}_{l'})^*\right) \\
    \label{eq:xiy}
     \xi^y &= i\sqrt{\eta_{l'}/2}\left(\iota^{k-1}_{l'}-(\iota^{k-1}_{l'})^*\right) \;,
\end{align}
and
\begin{align}
\label{eq:xiz}
    \xi^z= \frac{\eta_{l'} \sum_{r=1}^R \gamma_r^{k-1} \mu_{l'r}}{\sum_{r=1}^R (\mu_{l'r})^2 }\;.
\end{align}
The coefficients $\eta_{l'}$ and $\mu_{l'r}$ are 
previously known, and the coefficients $\gamma_r^{k-1}$
are also known from the $(k-1)-$th step.
Further details about Eq.~\eqref{eq:xiz} are discussed in Appendix~\ref{app:B}.
The unitary operation $V^\dagger_k$ is the one that diagonalizes
the Hamiltonian $\xi^z S^z_{l'} + \xi^x S^x_{l'}+ \xi^y S^y_{l'}$,
which appears on the right hand side of Eq.~\eqref{eq:kHamildecomp},
and transforms it to an observable that is proportional to $S^z_{l'}$.
Thus, $V_k$
is $e^{i(\pi^x S^x_{l'} + \pi^y S^y_{l'})}$, where the coefficients are
\begin{align}
\label{eq:pis}
    \pi^x=\theta\xi^y \; , \pi^y = -\theta \xi^x \; , 
\end{align}
and the rotation angle or phase is
\begin{align}
    \label{eq:theta}
    \theta=\arctan \left( \frac{\sqrt{(\xi^x)^2+(\xi^y)^2}}{\xi^z}\right) \;.
\end{align}
$V_k$ can be rewritten as
\begin{align}
    \exp \left\{ \frac i {\sqrt{2 \eta_{l'}}}\left( (\pi^x-i\pi^y)E^+_{l'} + (\pi^x+i\pi^y)E^-_{l'}\right)\right\}\;.
\end{align}
Then, the exponents of Eq.~\eqref{eq:GroupOps} obtained at the $k$-th
step of the diagonalization procedure are given by
\begin{align}
\label{eq:alphak}
    \alpha_k = \frac {\pi^x-i\pi^y}{\sqrt{2 \eta_{l'}}}\;,
\end{align}
and $l(k)=l'$ in Eq.~\eqref{eq:GroupOps}.
Using Eqs.~\eqref{eq:xix},~\eqref{eq:xiy},~\eqref{eq:xiz}, and~\eqref{eq:pis},
 $\alpha_k$ is completely determined by $\eta_{l'}$, $\iota^{k-1}_{l'}$,
and the coefficients $\gamma_r^{k-1}$ and $\mu_{l'r}$, for all $r \in \{1,\ldots, R\}$.
Last, we need
to obtain the new coefficients $\gamma_r^k$ and $\iota_l^k$
that will be used for the next step of the diagonalization procedure.
This can be done by working with a faithful representation
of $\mathfrak h$ of small dimension, as explained below.

In Ref.~\cite{Som05}, we show how $F_\psi^k=(V_k)^\dagger.F^{k-1}_\psi.V_k$
can be obtained by using the so-called adjoint representation of $\mathfrak h$.
This representation is faithful and is built upon $M\times M$-dimensional
matrices. Using the coefficients $\gamma_r^{k-1}$ and $\iota_l^{k-1}$,
we first construct the matrix $\bar F_\psi^{k-1}$ that represents $F_\psi^{k-1}$.
Next, using the coefficients 
$\alpha_k$, we construct $\bar V_k = e^{i(\alpha_k \bar E^+_{l'}
+\alpha^*_k \bar E^-_{l'})}$, which is an $M\times M$-dimensional
unitary matrix, and $\bar E^{\pm}_l$ are the representations of $E^\pm_l$.
We can compute $\bar F^{k}_\psi=(\bar V_k)^\dagger .\bar F^{k-1}_\psi .\bar V_k$
by simple matrix multiplication. 
The coefficients $\gamma_r^{k}$ and $\iota_l^{k}$
are proportional to $\Tr[\bar F^{k+1}_\psi. \bar H_r]$
and $\Tr[\bar F^{k+1}_\psi. \bar E^-_l]$, respectively.
This is because one can define an inner product
using the adjoint representation, where Eq.~\eqref{eq:hinnerproduct}
is satisfied if we replace $O_m \rightarrow \bar O_m$,
up to an irrelevant constant of proportionality.

We now analyze the overall
complexity of the diagonalization procedure.
To this end, we disregard roundoff errors
and obtain complexity estimates
by counting the number of simple operations,
such as summation, multiplication, and the computation
of simple trigonometric functions [e.g., $\arctan(.)$].

First, the $L$ $\mathfrak{su}(2)$
algebras are completely specified by 
the coefficients $\mu_{lr}$ and $\eta_l$, for all $l \in \{1 ,\ldots, L\}$
and $r \in \{1 ,\ldots, R\}$. We assume that the $\mu_{lr}$ 
can be directly obtained from the given commutation relations.
Since $L\le M$ and $R\le M$, the complexity associated with this step is $\cO(M^2)$.
This is basically the complexity of storing all these coefficients.
Additionally, each $\eta_l$ is very simple to obtain from the known commutators with 
complexity that does not dominate this step.

In the $k$-th step of the diagonalization procedure,
we assumed that the coefficients $\gamma_r^{k-1}$ and $\iota_l^{k-1}$
were also known and stored in a list. This assumption is valid for $k=1$,
where these coefficients are easily determined from the expectations 
$\langle O_m \rangle_\psi$.
Finding the $l'$ such that $|\iota^{k-1}_{l'}| \ge |\iota_l^{k-1}|$,
for all $l$, can be done with complexity $\tilde \cO(L)$ or $\tilde \cO(M)$.
Also, computing the coefficients $\xi^x$ and $\xi^y$ can be done
with complexity $\cO(1)$, and $\xi^z$ with complexity $\cO(R)$ or $\cO(M)$,
according to Eqs.~\eqref{eq:xix},~\eqref{eq:xiy}, and Eq.~\eqref{eq:xiz}.
It follows from Eq.~\eqref{eq:alphak} that $\alpha_k$ can also be obtained with complexity $\cO(M)$,
for given $l'$.

Once $\alpha_k$ is obtained, we need to 
update the coefficients as $\gamma_r^{k-1} \rightarrow \gamma_r^{k}$ and $\iota_l^{k-1}
\rightarrow \iota_l^{k}$, so that they can be used in the next step. 
This is the most expensive part
of the procedure. From the previous discussion,
the complexity of computing $\bar F_\psi^{k-1}$
is $\cO(M^3)$: a linear combination
of $M$ matrices of dimension $M\times M$.
Obtaining $\bar V_k$ can be done with complexity $\cO(M^3)$
using standard techniques for matrix exponentiation.
And obtaining $\bar F^{k}_\psi$ can also be done with complexity $\cO(M^3)$
via direct matrix multiplication
methods.
The updated coefficients $\gamma_r^{k}$ and $\iota_l^{k}$
require computing $\cO(M)$ traces of a product of two matrices,
and each trace can be obtained with complexity $\cO(M^2)$.
Thus, the overall complexity of updating all the coefficients
is $\cO(M^3)$, when using the adjoint representation.

In summary, the complexity of the diagonalization
procedure is dominated by transforming the Hamiltonian 
at each step (taking
$F_\psi^{k-1}\rightarrow F_\psi^k$), and this is $\cO(M^3)$. For $K'$ steps,
the overall complexity is $\cO(K'M^3)$.
This may be improved by using faster algorithms for 
various matrix
operations and by working with faithful representations of $\mathfrak h$
of smaller dimension, when possible.

In the previous analysis, we assumed that our starting Hamiltonian
was $F_\psi$ or, equivalently, we assumed that the expectation values
$\langle O_m \rangle_\psi$ were exactly known. In the actual scenario,
these expectations are known within certain precision and 
the starting Hamiltonian in the diagonalization
procedure is thus $\tilde F_\psi$, which is only an approximation
of $F_\psi$.

\subsection{Precision and confidence levels}
\label{sec:prec-conf}
The  described diagonalization procedure
is guaranteed to output a diagonal Hamiltonian
only when the number of steps, $K'$, 
is infinity. For $K'<\infty$,
the resulting Hamiltonian is only
almost diagonal, in the following sense.
We let $d^k=\sum_{l=1}^L |\iota_l^k|^2$
be the (squared) distance of $F_\psi^k$ to the CSA.
In particular, $d^k=0$ if and only if $F_\psi^k$
is diagonal, i.e., it belongs to the CSA.
Equation 5.63 of Ref.~\cite{Som05} implies that
\begin{align}
\label{eq:DPsteps}
    K' \ge \frac{\log(d^0/\epsilon_D)} {\log((L+1)/L)}
\end{align}
steps of the diagonalization procedure suffice to achieve
$d^{K'} \le \epsilon_D$.
That is, $K'=\cO(L \log(d^0/\epsilon_D))$, assuming $d^0 \ge \epsilon_D$.
Additionally, the expectation values $\langle O_m \rangle_\psi$
can only be estimated within certain precision and confidence levels
by finitely-many measurements. It follows that we can only 
provide a sequence of simple unitaries to prepare a quantum state
that is $\epsilon$-close to the GCS and
with confidence level $1-\delta$ -- see Sec.~\ref{sec:problem}.
For given $\epsilon$ and $\delta$, the complexity of the diagonalization
procedure and the number of measurements can be determined, as shown below.
 
The actual input Hamiltonian to the diagonalization procedure is
\begin{align}
    \tilde F_\psi = \sum_{m=1}^M O_m\tilde{\langle O_m \rangle}_\psi  \;.
\end{align} 
 Here, $\tilde{\langle O_m \rangle}_\psi$ are the estimates
 of the expectations, and we assume that each estimate is obtained
 within precision $\epsilon_M$. The Hamiltonian obtained at the $k-$th 
 step is $\tilde F_\psi^k =(\tilde V_k)^\dagger . \tilde F_\psi^{k-1}.\tilde V_k $, where $\tilde V_k$ is the Lie-group transformation discussed
 in Sec.~\ref{sec:DP}. 
 After $K'$ steps, the Hamiltonian is $\tilde F_\psi^{K'}$,
 and this may not be in the CSA. To this end, we regard
 as the output of the procedure the projection of $\tilde F_\psi^{K'}$
 into the CSA, that is, its diagonal part. We define it to be 
 $\tilde F_\psi^{\rm CSA}$.

 It is useful to define $\| O \|=\max_{1 \le m  \le M}\|O_m\|$ as the maximum 
of the operator norm of the observables in the orthogonal basis of
$\mathfrak h$.
We first set $\epsilon_D$ and $K'$. 
Under the assumption that $\epsilon_M \ll 1$,
the eigenvalues of $\tilde F_\psi$
are sufficiently close to those of $F_\psi$.
That is, the largest eigenvalue of $\tilde F_\psi$
is close to $P_{\mathfrak h}$ and the second 
largest one is close to $P_{\mathfrak h}-\Delta$, where $\Delta >0$ is the spectral gap of $F_\psi$. We will set a lower bound on $\Delta$ below.
In Appendix~\ref{app:A}, we use perturbation theory to show that there is
$\epsilon_D=\Omega(\epsilon^2 \Delta^2/(L \|O\|^2))$ such that
$\tilde F^{K'}_\psi$
is sufficiently close to being diagonal.
That is, if $\sket{\tilde{\rm w}_0}$ is the
eigenstate of largest eigenvalue of $\tilde F_\psi^{\rm CSA}$
and $\tilde V^\dagger$ is the unitary transformation to obtain
$\tilde F_\psi^{K'}$ from $\tilde F_\psi$, then
\begin{align}
\label{eq:GCSerror}
    \| \ket \psi- \tilde V  \sket{\tilde{\rm w}_0}  \|=\cO(\epsilon) \;,
\end{align}
in the limit $\epsilon_M \rightarrow 0$.
According to Eq.~\eqref{eq:DPsteps},
the number of steps is
\begin{align}
\label{eq:K'}
  K' = \cO \left ( L \log \left( \frac{d^0 L \|O\|^2}{\epsilon^2 \Delta^2}\right)\right) \;.
\end{align}
A sequence of Weyl reflections can transform $\ket{\tilde{\rm w}_0}$
into $\ket{\rm {hw}}$ as
\begin{align}
\label{eq:whwmap}
    U_{K'+1}\ldots U_K \ket{\rm {hw}} = \ket{\tilde{\rm {w}}_0}\;.
\end{align}

In Appendix~\ref{app:A}, we also show
that it suffices to have estimates of the expectations
within precision
$\epsilon_M=\Omega(\epsilon \Delta/(M \|O\|))$. (We assume, for simplicity,
that all estimates are obtained within the same precision.)
This choice of $\epsilon_M$ also implies Eq.~\eqref{eq:GCSerror}.

If we perform projective measurements
of each $O_m$ on $\ket \psi$, we obtain values
$O_m^i$ with probabilities $p_m^i$. 
According to Hoeffding's inequality~\cite{Hoe63}, after performing $Q$
projective measurements of $O_m$ to get the estimate
$\langle \tilde{O_m} \rangle_\psi$, we obtain
\begin{align}
    \Pr \left(\left | \langle \tilde{O_m} \rangle_\psi - {\langle O_m \rangle}_\psi \right| \ge \epsilon_M \right) \le 2 e^{-\frac{Q \epsilon_M^2}{2\|O\|^2}} \;.
\end{align}
We use a union bound for which, if the overall desired
confidence level is $1-\delta$, the confidence level in 
the estimation of each expectation value is, at least, $1-\delta/M$.
Under this assumption, it suffices
to have a number of measurements for each $O_m$ that satisfies
\begin{align}
\label{eq:PM}
    Q =\left \lceil \frac{2 \|O\|^2 \log(2M/\delta)}{\epsilon_M^2} \right \rceil \;.
\end{align}
The total number of measurements for all observables is $QM$.

Since $L =\cO( M)$, $d^0=\cO(M \|O\|^2)$, and $\epsilon_D=\Omega(\epsilon^2 \Delta^2/(M \|O\|^2))$,  Eq.~\eqref{eq:K'} is
\begin{align}
    K' = O \left( M  \log \left( \frac{M^2 \|O\|^4}{\epsilon^2 \Delta^2}\right) \right) \;.
\end{align}
The number of steps $K'$, in addition to the $\cO(M)$ Weyl reflections of Eq.~\eqref{eq:whwmap},
is the total number of unitaries $K$ of Eq.~\eqref{eq:ProcedureError}. 
The complexity of the diagonalization procedure, as discussed in Sec.~\ref{sec:DP},
is $\cO(K'M^3)$. This is 
\begin{align}
    \cO \left( M^4  \log \left( \frac{M^2 \|O\|^4}{\epsilon^2 \Delta^2}\right) \right) \;,
\end{align}
and thus polynomial in the dimension of $\mathfrak h$.
The number of projective measurements is
\begin{align}
   Q.M= \cO \left( \frac{M^3 \|O\|^4 \log(M/\delta)}{\epsilon^2 \Delta^2}\right) \;,
\end{align}
which was obtained using Eq.~\eqref{eq:PM} and $\epsilon_M=\Omega(\epsilon \Delta/(M \|O\|))$.
This number is also polynomial in $M$.

Last, we make a remark about the spectral gap
$\Delta$, which is also the spectral gap of $F_{\rm hw}$.
By construction, $\Delta$ scales as $\|O\|^2$ in the sense that
$\Delta \rightarrow \kappa^2 \Delta$ if we replace $O_m \rightarrow \kappa O_m$,
$\kappa >0$.
$\Delta$ can be determined from the commutation
relations of $\mathfrak h$ or, more precisely, by the structure
constants of the algebra. This is because other weight states can be obtained
from $\ket{\rm{hw}}$ by acting with the lowering operators, and the structure
constants can be used to determine the weights and eigenvalues of the
observables in the CSA.
As these structure constants are, a priori, independent of
the dimension $M$, the second largest eigenvalue of $F_{\rm hw}$ has to be
at a gap of order $c \|O\|^2$ from the largest eigenvalue, with $c=\cO(1)$. 
Then,
the complexity of the diagonalization procedure is $\cO(M^4 \log(M/\epsilon))$
and the number of projective measurements is $\cO(M^3 \log(M/\delta)/\epsilon^2)$.

The unitaries of Eq.~\eqref{eq:whwmap} together with the $K'$ unitaries $U_k:=\tilde V_k$,
 which result from the diagonalization procedure applied to $\tilde F_\psi$, provide the 
 $K$ unitaries that satisfy Eq.~\eqref{eq:ProcedureError}.

\section{Applications to the simulation of quantum circuits}
\label{sec:applications}
In Ref.~\cite{SBO06} we introduced the
model of Lie-algebraic quantum computing (LQC).
In this model, the set of gates is induced 
by the control Hamiltonians that belong to a semisimple Lie algebra $\mathfrak h$,
and that set may not be universal. The initial state
is a ground state of a control Hamiltonian  (e.g., a highest-weight state)
and the final measurement is an expectation of either a control Hamiltonian 
or a unitary operator in the group generated by $\mathfrak h$.
A main result of Ref.~\cite{SBO06} is 
that an LQC computation can be simulated classically efficiently in the dimension of $\mathfrak h$; in some cases, this is a significant improvement over the obvious strategy whose complexity must be at least linear in the dimension of the Hilbert space.

The results of this paper can then be used to obtain a bigger
 class of quantum circuits that can be classically simulated.
 In more detail, let a quantum circuit be described by the unitary
 $T=T_1.T_2\ldots T_\cL$, where each $T_l$ is also unitary. 
 The main variant with respect to LQC
 is that the unitaries may not be induced by the observables in $\mathfrak h$.
 However, we still assume that the quantum state
 prepared after the action of each $T_l$ is a GCS. We also require that the transformed observable $T^\dagger_l O_m T_l$,
 which can be decomposed as a linear combination of observables acting on $\cH$, can be obtained efficiently. More specifically, we assume
 that $\Tr[O_{m'}(T^\dagger_l O_m T_l)]$ can be computed classically in time
 that is polynomial in $M$, for all $l \in \{1,\ldots,\cL\}$
 and $m,m' \in \{1,\ldots,M\}$. We note that these two assumptions
 are satisfied if the $T_l$'s belong to the Lie group induced by $\mathfrak h$,
 as in LQC.
 
 Let $\ket{\psi_l}$ be the GCS after the action of $T_l$ and
 $\ket{\psi_0}$ be the initial state.
 As described in Sec.~\ref{sec:GCS}, $\ket{\psi_l}$ is completely
 specified (up to an irrelevant phase) by the expectation values
 $\langle O_m \rangle_{\psi_l}$. We also note that
 \begin{align}
     \langle O_m \rangle_{\psi_{l}}= \langle T^\dagger_l O_m T_l \rangle_{\psi_{l-1}} \;.
 \end{align}
 Thus, the expectation values after the action of $T_l$
 can be obtained from the previous ones in time that is polynomial in $M$,
 under the assumptions. In particular, we can compute
 the expectation values $\langle O_m \rangle_{\psi_{\cL}}$ of the final
 state on a classical computer
 in time that is polynomial in $M$ and linear in the length of the
 quantum circuit, $\cL$.
 
 If the goal is to obtain
 an expectation value of an observable in $\mathfrak h$ on $\ket{\psi_\cL}$,
 such an expectation can be readily obtained from the computed
 $\langle O_m \rangle_{\psi_{\cL}}$'s. If, however, the goal
 is to obtain the expectation value of a unitary operator in the 
 group or an observable that is not in $\mathfrak h$, we can proceed as follows.
 First, from the  $\langle O_m \rangle_{\psi_{\cL}}$'s
 we build the Hamiltonian
 \begin{align}
     F_{\psi_\cL}=\sum_{m=1}^M O_m \langle O_m \rangle_{\psi_{\cL}} \;.
 \end{align}
 Then, we run the diagonalization procedure described in 
 Sec.~\ref{sec:DP} to obtain a sequence of Lie-group operations
 that prepare $\ket{\psi_l}$ from the highest-weight state.
 Once such a sequence is obtained, we have reduced the problem
 to that of simulating LQC, which can be done efficiently using classical resources
 according to the results in Ref.~\cite{SBO06}.
 
 Note that, for precision $\epsilon$ in the computation
 of the expectation value, the number of Lie-group
 unitaries obtained by the diagonalization procedure is $\cO(M \log(M/\epsilon))$. The complexity for implementing the diagonalization
 procedure is $\cO(M^4 \log(M/\epsilon))$. 
 Then, the overall complexity for classically
 simulating the quantum computation in this case is efficient; that is, polynomial in $M$
 and linear in $\cL$.

 \section*{Acknowledgements}
 We acknowledge support from the LDRD program
 at Los Alamos National Laboratory.

%

\begin{appendix}
\section{Coefficients for $\mathfrak{su}(2)$ operations}
\label{app:B}
As pointed out in Eqs.~\eqref{eq:Fdecomp} and~\eqref{eq:kHamildecomp}, we can write
\begin{align}
\label{eq:FdecompCSA}
    F_\psi^{k-1} &= \sum_{r=1}^R \gamma_r^{k-1} H_r + \ldots \\
\label{eq:Fdecompsu2}
    & = \xi^z S_{l'}^z + \ldots \;,
\end{align}
where the coefficients $\gamma_r^{k-1}$ are stored in memory.
The remaining terms in the right hand side of Eq.~\eqref{eq:FdecompCSA}
are in $\mathfrak h ^+ \oplus \mathfrak h^-$. The remaining terms in the right hand side of Eq.~\eqref{eq:Fdecompsu2} are linear combinations of $S_{l'}^{x}$ and $S_{l'}^{y}$,
and other observables orthogonal to the corresponding $\mathfrak{su}(2)$ algebra.

We seek to find the coefficient $\xi^z$ of Eq.~\eqref{eq:kHamildecomp}.
Using Eq.~\eqref{eq:hinnerproduct} and $Z_{l'}=\sum_{r=1}^R \mu_{l'r}H_r$, on one hand we have
\begin{align}
    \Tr[F_\psi^{k-1} . Z_{l'}]={\cal N} \sum_{r=1}^R \gamma_r^{k-1} \mu_{l'r} \;.
\end{align}
On the other hand, since $S^z_{l'}=Z_{l'}/\eta_{l'}$, we have
\begin{align}
    \Tr[F_\psi^{k-1} . Z_{l'}]&= (\xi^z/\eta_{l'})  \Tr[(Z_{l'})^2] \\
    \nonumber
    & ={\cal N} (\xi^z/\eta_{l'})  \sum_{r=1}^R (\mu_{l'r})^2 \;.
\end{align}
Then,
\begin{align}
    \xi^z= \frac{\eta_{l'} \sum_{r=1}^R \gamma_r^{k-1} \mu_{l'r}}{\sum_{r=1}^R (\mu_{l'r})^2 }\;,
\end{align}
which coincides with Eq.~\eqref{eq:xiz}.

\section{Approximation errors and perturbation theory}
\label{app:A}
Given a Cartan-Weyl basis, the
observables that determine an orthogonal basis for $\mathfrak h$
are given by Eq.~\eqref{eq:CWbasis}.
Then,
$\|E^\pm_l\| \le \|O\|$ for all $l\in \{1,\ldots,L\}$. 
When the input to the diagonalization procedure is $F_\psi$,
the coefficients
of the output Hamiltonian, after $K'$ steps, satisfy $\sum_{l=1}^L |\iota_l^{K'}|^2 \le \epsilon_D$.
Also, the term of $F_\psi^{K'}$ that
does not belong to the CSA has an operator norm that is
upper bounded by
\begin{align}
\label{eq:appbound1}
   2 \|O\| \sum_{l=1}^L |\iota_l^{K'}|  \le 2  \|O\| \sqrt{\epsilon_D L} \;,
\end{align}
where the last inequality follows from a property between the 1- and 2-norm of complex
vectors.

The Hamiltonian $F_\psi^{K'}$ can be written as
\begin{align}
  F_\psi^{K'} = F_\psi^{\rm{CSA}}  + F_\psi^{\perp} \;,
\end{align}
where $F_\psi^{\rm{CSA}} \in \mathfrak h_D$ and
$F_\psi^{\perp}$ is orthogonal to the CSA in the sense of
Eq.~\eqref{eq:hinnerproduct}; that is $F_\psi^{\perp} \in \mathfrak h^+ \oplus \mathfrak h^-$. Equation~\eqref{eq:appbound1}
implies
\begin{align}
\label{eq:appbound2}
    \| F_\psi^{\perp} \| = {\cO} \left(  \|O\| \sqrt{\epsilon_D L} \right)\;.
\end{align}
We let $\ket{\rm w_0}$ be the eigenstate (weight state) of largest eigenvalue
of $F_\psi^{\rm{CSA}}$ and $\ket{ {\rm w}_1}$
be the eigenstate of largest eigenvalue of $F_\psi^{K'}$. Note that
$\ket{ {\rm w}_1}$ is not, in principle, a weight state.
We seek an expression for an upper bound on $\|\ket{ {\rm w}_1} - \ket{ {\rm w}_0} \|$, which can be obtained from perturbation theory. 
$\epsilon_D$ will then be obtained by requiring such a bound to be
$\cO(\epsilon)$.

To obtain this bound, we parametrize the Hamiltonian
as
\begin{align}
    F_\psi^{K'} (s)= F_\psi^{\rm{CSA}}  +s F_\psi^{\perp}\;,
\end{align}
with $0 \le s \le 1$. The eigenstate of largest eigenvalue is defined to be $\ket{{\rm w}_s}$. 
In the perturbative
regime, where the dominant term in the Hamiltonian is
$F_\psi^{\rm{CSA}}$, $\ket{{\rm w}_s}$ determines a differentiable path in Hilbert space.
Then,
\begin{align}
\nonumber
    \|\ket{ {\rm w}_1} - \ket{ {\rm w}_0} \| &= 
    \left \| \int_0^1 ds \ket{\partial_s {\rm w}_s } \right \|
    \\
    \nonumber
    & \le \int_0^1 ds \| \ket{\partial_s {\rm w}_s } \|
    \\
    \label{eq:appboundstate}
    & \le \max_{0 \le s \le 1} \| \ket{\partial_s {\rm w}_s } \|\;.
\end{align}

We can use the eigenvalue equation
$F_\psi^{K'} (s) \ket{{\rm w}_s} =
 \lambda(s) \ket{{\rm w}_s}$ to obtain
\begin{align}
    \ket{\partial_s {\rm w}_s } = \frac 1 {\lambda(s) -F_\psi^{K'} (s) }(F_\psi^\perp - \partial_s \lambda(s))\ket{ {\rm w}_s } \;,
\end{align}
and then
\begin{align}
\label{eq:appbound6}
\| \ket{\partial_s {\rm w}_s }\| \le \max_{0 \le s \le 1}
\frac {2 \|F_\psi^\perp\|} {\Delta(s)} \;.
\end{align}
Here, $\Delta(s)$ is the spectral gap between the largest
and second largest eigenvalues of $F_\psi^{K'}(s)$. In particular,
in the perturbative limit where $\|F_\psi^\perp\| \le \Delta/4$,
$\Delta(s) \ge \Delta/2$ as each eigenvalue of $F_\psi^{K'}(s)$ differs
from the eigenvalues of $F_\psi^{K'}$ by, at most, $\Delta/4$.
Then, $\Delta(s)=\Omega(\Delta)$. To obtain Eq.~\eqref{eq:appbound6},
we also used the property $\partial_s \lambda(s)= \bra{ {\rm w}_s } F_\psi^\perp \ket{ {\rm w}_s }$. Choosing $\epsilon_D=\cO(\epsilon^2 \Delta^2/(L \|O\|^2))$,
Eqs.~\eqref{eq:appbound2},~\eqref{eq:appbound6}, and~\eqref{eq:appboundstate} imply
\begin{align}
    \label{eq:appbound7}
    \|\ket{ {\rm w}_1} - \ket{ {\rm w}_0} \| = \cO \left(\epsilon \right)\;.
\end{align}
Our choice of $\epsilon_D$ implies $\|F_\psi^\perp \|=\cO(\epsilon \Delta)$
in Eq.~\eqref{eq:appbound2}.
Note that $\|F_\psi^\perp \| \le \Delta/4$ for sufficiently small $\epsilon$,
justifying the arguments about the perturbative limit.

Equation~\eqref{eq:appbound7} implies that, if $V^\dagger$
is the unitary Lie-group transformation that mapped $F_\psi$ to
$F_\psi^{K'}$ via the diagonalization procedure, i.e. $F_\psi^{K'}=V^\dagger.
F_\psi. V$,
then
\begin{align}
\label{eq:appbound10}
    \| \ket \psi - V \ket{\rm w_0} \| = \cO(\epsilon) \;.
\end{align}
This implies that, if the expectation values are known exactly,
the diagonalization procedure can be used to obtain a sequence
of Lie-group operations that transform a weight state $\ket{\rm w_0}$
to a state that is $\cO(\epsilon)$-close to the GCS $\ket \psi$.

The weight state $\ket{\rm w_0}$ can be transformed
to the highest-weight state by a sequence of Lie-group
operations referred to as Weyl reflections. To show this,
we simply need to show that both, $\ket{\rm w_0}$ and $\ket{\rm {hw}}$
belong to the orbit of GCSs and thus are connected via a Lie-group transformation.
The proof is by contradiction: If $\ket{\rm w_0}$ is not a GCS, then
$V\ket{\rm w_0}$ is orthogonal to $\ket \psi$ and Eq.~\eqref{eq:appbound10}
is not satisfied (assuming small $\epsilon$). These Weyl reflections, in addition to the $K'$ unitaries
output by the diagonalization procedure, provide a sequence
of $K$ simple Lie-group operations to transform $\ket{\rm hw}$
into a state that is $\cO(\epsilon)$-close to $\ket \psi$.

The effects of additional errors from not knowing the expectation values exactly
are now analyzed.
In the actual case, the starting Hamiltonian is
\begin{align}
    \tilde F_\psi = \sum_{m=1}^M O_m \langle \tilde{O_m}\rangle_\psi \;,
\end{align}
where $\langle \tilde{O_m}\rangle_\psi$ are estimates of 
$\langle {O_m}\rangle_\psi$ within precision $\epsilon_M$.
(We assume, for simplicity, that all the expectations 
are known within the same precision.)
It follows that
\begin{align}
\label{eq:appbound8}
    \left \|   F_\psi - \tilde F_\psi \right \| = \cO \left( M \|O\| \epsilon_M \right) \;.
\end{align}
We will use perturbation theory to make several useful claims about
$\tilde F_\psi$.

Let $\sket{\tilde \psi}$ be the eigenstate of largest eigenvalue
of $\tilde F_\psi$, which is also a GCS. 
A similar analysis as the one that resulted in Eq.~\eqref{eq:appbound6}
for this case
implies 
\begin{align}
    \| \ket \psi  - \sket{\tilde \psi} \|\le \frac{2 \|   F_\psi - \tilde F_\psi  \| }{ \Delta'} \;,
\end{align}
where $\Delta'$ is the minimum spectral gap of the Hamiltonian
that continuously interpolates between $F_\psi$ and $\tilde F_\psi$.
In the perturbative regime where $\|F_\psi -\tilde F_\psi\| =\cO( \epsilon \Delta)$, $\Delta'=\Omega(\Delta)$
and 
\begin{align}
\label{eq:appboundstate2}
    \|\ket \psi  - \sket{\tilde \psi} \| = \cO(\epsilon)\;.
\end{align}

The diagonalization procedure will then be applied
to $\tilde F_{\psi}$. To satisfy Eq.~\eqref{eq:appboundstate2},
it then suffices to set $\epsilon_M = \cO(\epsilon \Delta/(M\|O\|))$
in Eq.~\eqref{eq:appbound8}.
The largest and second largest eigenvalues of $\tilde F_{\psi}$
differ from those of $F_{\psi}$ by an amount of order $\epsilon \Delta$.
Then, the spectral gap of $\tilde F_{\psi}$ is also $\Omega(\Delta)$ and
a perturbative analysis also implies that $\epsilon_D=\cO(\epsilon^2 \Delta^2/(L \|O\|^2))$ suffices to output a Hamiltonian that is sufficiently close
to being diagonal. 

Our choice of $\epsilon_D$ determines
the number of steps $K'$ for the diagonalization procedure.
Let the Hamiltonian at the $K'$-th step
be $\tilde F_\psi^{K'}=\tilde F_\psi^{\rm{CSA}}+
\tilde F_\psi^\perp$, where $\tilde F_\psi^{\rm{CSA}}\in \mathfrak h_D$
and $\tilde F_\psi^\perp \in \mathfrak h^+ \oplus \mathfrak h^-$.
We also let $\sket{\tilde {\rm w}_1}$ and $\sket{\tilde {\rm w}_0}$
be the eigenstates of largest eigenvalue of $\tilde F_\psi^{K'}$
and $\tilde F_\psi^{\rm{CSA}}$, respectively. Following a similar analysis
as the one that led to Eq.~\eqref{eq:appbound7}, we obtain
\begin{align}
    \label{eq:appbound20}
    \|\sket{ \tilde{\rm w}_1} - \sket{ \tilde{\rm w}_0} \| = \cO \left(\epsilon \right)\;.
\end{align}

The desired output of the diagonalization procedure 
is then $\tilde F_\psi^{\rm{CSA}}$ and the relevant 
eigenstate is the weight state $\sket{ \tilde{\rm w}_0}$.
If $\tilde V^\dagger$ is the unitary Lie-group transformation
that satisfies $\tilde F^{K'}_\psi=\tilde V^\dagger. \tilde F_\psi. \tilde V$,
which was obtained via the diagonalization procedure, then
\begin{align}
\nonumber
    \|\ket \psi  -  \tilde V \sket{ \tilde{\rm w}_0}\| &\le
    \|\ket { \psi}  -\sket {\tilde \psi} \| +
    \| \sket {\tilde \psi} - \tilde V \sket{ \tilde{\rm w}_1}\| + \\
    \label{eq:appboundstate3}
    & \ \ + \|\tilde V \sket{ \tilde{\rm w}_1}  - \tilde V \sket{ \tilde{\rm w}_0}\|
     \;.
\end{align}
Equation~\eqref{eq:appboundstate2}, our choice of $\epsilon_D$, and Eq.~\eqref{eq:appbound20}, imply that each term
in the right hand side of Eq.~\eqref{eq:appboundstate3} is $\cO(\epsilon)$,
respectively. In particular,
\begin{align}
     \|\ket \psi  -\tilde V \sket{ \tilde{\rm w}_0}  \| = \cO(\epsilon)\;.
\end{align}

In summary, when the input to the diagonalization
procedure is the Hamiltonian $\tilde F_\psi$ obtained
from an estimate of the expectation values, the output
is a Hamiltonian $\tilde F_\psi^{\rm{CSA}}$ and a Lie-group operation
$\tilde V$, given as a sequence of exponentials of the form
of Eq.~\eqref{eq:GroupOps}. The eigenstate
of largest eigenvalue of $\tilde F_\psi^{\rm{CSA}}$
is a weight state $\sket{ \tilde{\rm w}_0}$, and if acted on
by $\tilde V$, it outputs a GCS that is $\cO(\epsilon)$-close to $\ket \psi$.
A sequence of simple Weyl reflections transforms $\sket{ \tilde{\rm w}_0}$
into the highest-weight state. These reflections, together with the unitaries
obtained by the diagonalization procedure, provide
the unitaries that satisfy Eq.~\eqref{eq:ProcedureError}.

\end{appendix}

\end{document}